\documentclass[review]{elsarticle}
\usepackage{amsmath,amssymb,amsthm,amsfonts}
\usepackage{epsfig}
\usepackage{subfigure}
\usepackage{enumerate}
\usepackage{color}
\usepackage{float}
\usepackage{graphicx}
\newfont{\tenbfit}{cmmib10}%
\newfont{\svnbfit}{cmmib8}%
\newfont{\tenbfsl}{cmbxti10}% <-- for idem tensor
\newfont{\mmit}{cmmi10}% scaled 1200
\newfont{\smit}{cmmi9}%  scaled 1200
\newfont{\bfMit}{cmmi5}%  scaled 1200
\newfont{\tenbbb}{msbm10}%
\newfont{\svnbbb}{msbm8}%
\newfont{\tenssit}{cmssqi8 at 10pt}%
\newfont{\svnssit}{cmssqi8 at 7pt}%
\newfont{\gothic}{eufm10}%
\newfont{\sgothic}{eufm7}%

\newcommand{\pards}[2]{\mbox{$\dfrac{\partial #1}{\partial {#2 }}$}}

\newcommand{\Blj}{\mbox{$\Big[\kern-0.275em\Big[$}}
\newcommand{\Brj}{\mbox{$\Big]\kern-0.275em\Big]$}}

\theoremstyle{plain}
\usepackage{titleref}
\usepackage{xr}

\journal{Journal of Mechanics and Physics of Solids}

\makeatletter
\def\ps@pprintTitle{%
   \let\@oddhead\@empty
   \let\@evenhead\@empty
   \let\@oddfoot\@empty
   \let\@evenfoot\@oddfoot
}
\makeatother

\begin{document}
\begin{frontmatter}
 \title{Effect of Elasticity on Phase Separation in Heterogeneous Systems}

\author[CEE]{Mrityunjay Kothari}
\author[CEE,ME]{Tal Cohen\corref{mycorrespondingauthor}}

\cortext[mycorrespondingauthor]{Corresponding author}
\ead{talco@mit.edu}

\address[CEE]{Department of Civil and Environmental Engineering, Massachusetts Institute of Technology, Cambridge, MA, USA, 02139}
\address[ME]{Department of Mechanical Engineering, Massachusetts Institute of Technology, Cambridge, MA, USA, 02139}

\begin{abstract}
A recent study has demonstrated that phase separation in binary liquid mixtures is arrested in the presence of elastic networks and can lead to a nearly uniformly-sized distribution of the dilute-phase droplets. At longer timescales, these droplets exhibit a directional preference to migrate along elastic property gradients to form a front of dissolving droplets [K. A. Rosowski, T. Sai, E. Vidal-Henriquez, D. Zwicker, R. W. Style, E. R. Dufresne, Elastic ripening and inhibition of liquid-€"liquid phase separation, Nature Physics (2020) 1-€"4]. In this work, we develop a complete theoretical understanding of this phenomenon in nonlinear elastic solids by employing an energy-based approach that captures the process at both short and long timescales to determine the constitutive sensitivities and the dynamics of the resulting front propagation. We quantify the thermodynamic driving forces to identify diffusion-limited and dissolution-limited regimes in front propagation. We show that changes in elastic properties have a nonlinear effect on the process. This strong influence  can have implications in a variety of material systems including food, metals, and  aquatic sediments, and further substantiates the hypothesis that biological systems exploit such mechanisms to regulate important function. 
\end{abstract}


\end{frontmatter}

\section{Introduction}
Phase separation is observed across the scales and in a variety of physical processes that range from the fields of biology to metallurgy. Liquid-liquid phase separation and subsequent species migration is understood to be a crucial mechanism for regulating the normal bodily functions of a cell. For instance, P-granule assembly is believed to employ phase separation as a mechanism to polarize the cell along its anterior-posterior axis as a precursor to cell differentiation \cite{brangwynne2009germline}. The dynamics of this process is dependent on the cytoplasm and its mechanical properties. In metals, phase transitions are extensively employed to tailor the mechanical properties for a variety of applications ranging from cryogenic temperature applications such as LNG transport \cite{kothari2019thermo} to high temperature applications like furnaces \cite{tancret2018phase, smith2016phase}. The morphology and distribution of precipitates resulting from phase transition in such metals is also affected by the presence of elastic fields \cite{nabarro1940influence, doi1985effects,fratzl1999modeling,karpov1998suppression}. This is of particular importance in the food industry, where phase separation controls how different components of the food interact, resulting in the texture and the flavor of the dish. The stable distribution of particles of one phase in another, such as water-in-oil in salad dressing, ice-in-cream in ice cream and air-in-protein network in bread not only impacts its flavor but also its longevity. Elasticity is also believed to play an important role in determining the size of gas bubbles that are formed in sediments, and their ability to migrate. For example, the growth of methane bubbles in aquatic sediments, and the subsequent transport of this potent greenhouse gas to the atmosphere can be significantly affected by the heterogeneous  elastic properties of the surrounding medium  \cite{johnson2002mechanical, algar2010stability, liu2018methane}. From these examples spanning various fields, we see that elasticity emerges as a common denominator that has a significant bearing on the final outcome of phase separation. Elasticity provides a way to not only deepen our understanding of biological principles of organisation but also to improve existing technologies to better control material microstructure and to improve the stability and shelf life of food. While important work has been carried out, both experimentally and theoretically, in understanding the elastic effects on phase separation and species migration in soft-material systems, a complete theoretical understanding of the dynamics of these processes still presents many challenges. In practical applications, the matrix properties are often heterogeneous, and the phase separation process is sensitive to these non-uniformities. Additionally, phase separation can induce large deformations in the elastic matrix, thus giving rise to nonlinear  material response, which can lead to distinct phase separation behavior even among materials with similar  properties at the linear elastic limit. Furthermore, the connection between elasticity and the timescales associated with the phase separation is not well understood. In this work we overcome these challenges to obtain a quantitative understanding of the role of elasticity in the short and long-time dynamics of phase separation. 

The interruption of phase separation due to the elastic resistance of a cross-linked network was first theoretically predicted by de Gennes \cite{de1979effect} and termed as \textit{microphase separation}. Subsequent experiments by Briber and Bauer \cite{briber1988effect} confirmed this microphase separation by considering the nucleation behavior in deuteriated polystyrene(PSD) and poly(vinyl methyl ether) (PVME), although the results showed deviation from the predicted trends. Recent works \cite{style2018liquid, rosowski2020elastic, vidal2020theory} have studied phase separation in a quenched mixture of liquid polydimethylsiloxane (PDMS) and fluorinated oil in crosslinked PDMS. The minority component (fluorinated oil) separated into nearly-monodisperse spherical droplets, spread throughout the crosslinked PDMS. It was reported that the stiffness of the polymer controls the size of the droplets and the nucleation temperature. However, the theoretical underpinnings of how elasticity controls the nucleation temperature, selects a particular equilibrium droplet size and can lead to the migration of species across stiffness gradients remain to be quantitatively understood.

There are two important timescales in the phase separation response of a liquid mixture in presence of elastic effects. At the short timescale, a rapid quench leads to the nucleation of liquid droplets of the dilute phase in the matrix. The droplets grow to a stable size dictated by the local elastic properties of the matrix. Any spatial heterogeneities in the stiffness of the matrix result in non-uniform solubility of the dilute phase. At the long timescale, these non-uniformities in the matrix concentration set up a diffusive flux in the matrix which alters the droplet sizes and can create a propagating front of dissolving droplets. In this paper, we first develop a theory to model the quasi-equilibrium phase separation response at the short timescale. This sets up the groundwork for modeling the long-time kinetics in spatially heterogeneous material systems and leads to a quantitative understanding of how elasticity controls the speed of the front. To validate our theory, we compare the predictions of our model with experimental results in the literature. In particular, we explain through our model the observation of a propagating droplet dissolution front in a composite elastic sample made of stiff and soft materials \cite{rosowski2020elastic}. 

\section*{Results}
\subsection*{\textbf{Theory of Phase Separation with Elastic Effects}}\label{section_2.1}%
\begin{figure}[H]
	\centering
  \includegraphics[scale = 0.5]{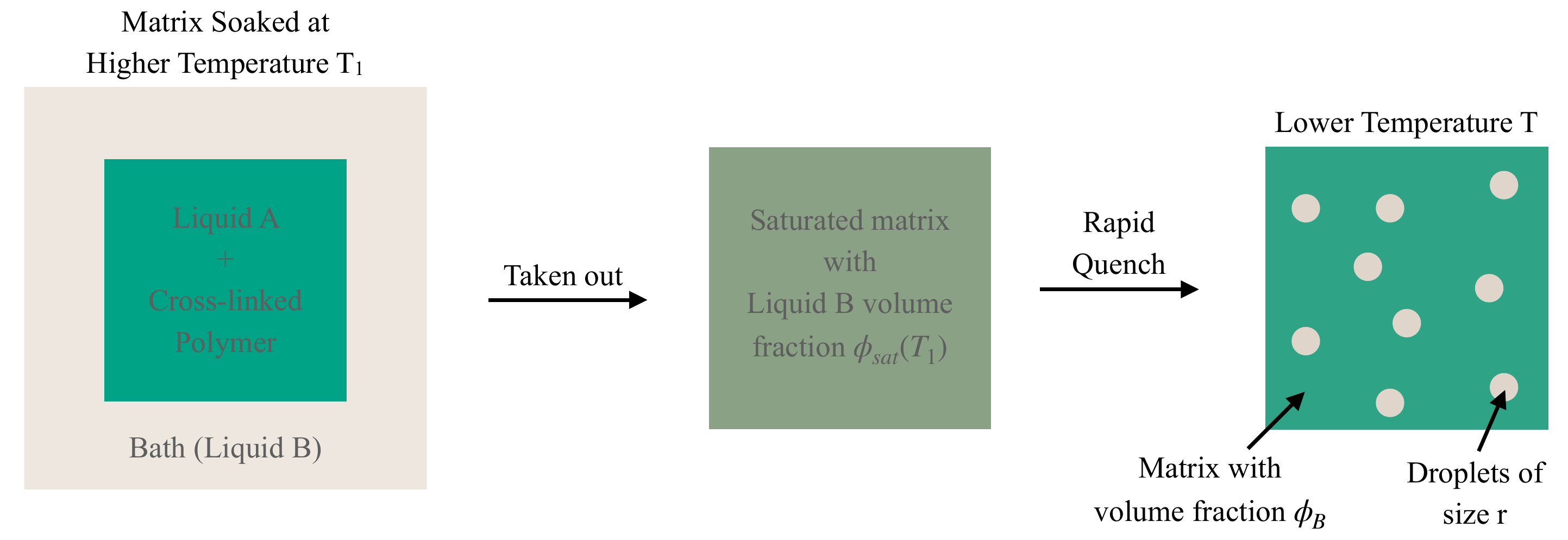}
  \caption{\textbf{Schematic of the phase separation process.} The matrix material made of liquid $A$ and crosslinked polymer, is soaked in a bath of liquid $B$ at an elevated temperature and saturated with a volume fraction $\phi_{sat}(T_1)$ of liquid $B$. Then, this assembly is quenched rapidly. The mixture phase separates into two components - (1) matrix consisting of crosslinked  polymer and liquids $A$ and $B$, and (2) droplets of liquid $B$. The size of droplets and the volume fractions are controlled by elasticity. }\label{fig1}
\end{figure}

\noindent Our model system consists of a partially cross-linked polymer with liquid $A$ that is soaked in a bath of liquid $B$ at temperature $T_1$. At equilibrium, this results in the polymer being saturated with a small volume fraction of liquid $B$, denoted by $\phi_{sat}(T_1)$. This assembly is taken out of the bath and rapidly quenched to a lower temperature $T$ (Fig. \ref{fig1}) where the mixture becomes unstable and demixes, giving rise to droplets of liquid $B$. We distinguish between two regions in our system; the droplets, which contain only liquid $B$, and the `matrix' that is occupied by the crosslinked network and permeated by liquids $A$ and $B$. 

For a  mathematical description of the system we introduce the following notations: $\phi_A=V_A/V_M$ and $\phi_B=V_B/V_M$ denote the volume fractions of liquids $A$ and $B$ in the matrix, where $V_M$ denotes the volume of the matrix within a representative unit element of the material system, and $V_A$, $V_B$ are the separate volumes of liquids $A$ and $B$ within the matrix, respectively. The total volume of the unit is $V=V_M+V_D$, where $V_D$ is the total volume of droplets. The volume fraction of droplets is  $\phi_D=V_D/V$. The initial volume of polymer within the unit element (without liquid $B$) can be written as $V_0=(1-\phi_B)V_M$, and the volume fractions of the two liquids are  related by \begin{equation}\label{pApB}
    \phi_A=\frac{V_A}{V_0}(1-\phi_B).
\end{equation} 

We make the following assumptions in our model. We consider the model in the thermodynamic size limit so that there are no effects due to the finiteness of the system. Liquid $B$ is the minority phase and has a small volume fraction ($\phi_B\ll 1$). Liquid $B$ is the only mobile component in the system; both $V_A$ and $V_0$ remain constant throughout the process. Due to the crosslinking, the polymer network acts as one large molecule which does not contribute to the mixing energy. Further, motivated by a recent experimental study \cite{style2018liquid}, we will take the droplet size distribution to be monodisperse, hence we write
\begin{equation}\label{pD_def}
\phi_D\equiv \frac{4\pi}{3} r^3n_d
\end{equation}
where $r$ denotes the radius of the droplets, and $n_d$ denotes their number density. We assume that the number density is determined by the initial kinetics of phase separation. Hence, we do not attempt to model these kinetics in this paper; guided by \cite{style2018liquid} we choose a linear dependence of $n_d$ on the matrix stiffness.

 The change in total free energy density of the system, calculated per unit  volume, can be expressed as the sum  of contributions from mixing ($\Delta \bar{G}_{mix}$), elastic ($\Delta \bar{G}_{el}$) and surface energies ($\Delta \bar{G}_{sur}$) as,
\begin{align} 
    &\Delta \bar{G}(\phi_A, \phi_B,\phi_D,T) = \Delta \bar{G}_{mix}(\phi_A, \phi_B,T) + \Delta \bar{G}_{sur}(\phi_D) +\Delta \bar{G}_{el}(\phi_D) \label{energy1}  \\ 
    &\Delta \bar{G}_{el}(\phi_D)=\frac{4\pi}{3}r^3W(r) n_d  \label{Gel} \\
    &\Delta \bar{G}_{sur}(\phi_D)= 4\pi r^2\Gamma n_d \label{Gsur}
\end{align}
where $W(r)$ is the strain energy stored in the elastic matrix per droplet and $\Gamma$ is the surface energy between the two liquids. Here the free energy density is written as a function of the volume fractions, such that $\Delta \bar{G}=\Delta \bar{G}(\phi_A, \phi_B,\phi_D,T)$. Using the above definitions, the free energy of a unit element (of volume $V$) can be written in terms of volumes $\Delta {G}=\Delta {G}(V_A, V_B,V_D,T)$.  In the following derivations we alternate between these representations for analytical convenience and clarity. 

Throughout the quenching and droplet growth process we assume a closed system; sum total of the volume of liquid $B$ dissolved in the matrix and in droplets must be the same as the initial saturation volume of liquid $B$ at $T_1$ before quenching,
\begin{equation} \label{mass_conservation}
    \phi_{B}(1-\phi_D) + \phi_D = \phi_{sat}(T_{1})
\end{equation}
Equations (\ref{energy1}-\ref{mass_conservation}) complete the system of equations to study the effect of elasticity on phase separation.

\subsection*{\textbf{Short Timescale: Elasticity Delays Phase Separation and Arrests Ostwald Ripening }}\label{section_2.2}%

The solubility of liquid $B$ in the matrix typically decreases with temperature \cite{style2018liquid, rosowski2020elastic,porter2009phase}. As the system is rapidly quenched from initial temperature $T_1$, it becomes supersaturated with liquid $B$. \noindent We define a critical temperature $T_c$ at which a phase-separated state containing $r_c$ sized droplets becomes lower in energy than the homogeneously mixed state. Thus, $T_c$ is given as the solution of the following equation,
\begin{equation} \label{nucleation}
    \Delta\bar{G}(\phi_A, \phi_{B,c}, \ \phi_{D,c} \ ,T) - \Delta\bar{G}(\phi_A, \phi_{sat}(T_1),0 ,T)=0
\end{equation}
where from \eqref{pD_def} and \eqref{mass_conservation} we have
\begin{equation}
    \phi_{D,c} = \frac{4\pi}{3}r_c^3 n_d \quad \text{and} \quad \phi_{B,c} = \frac{\phi_{sat}(T_1)-\phi_{D,c}}{1-\phi_{D,c}}.
\end{equation}

After the initial nucleation and growth of droplets, phase-separated liquid mixtures coarsen over time by growing larger droplets at the cost of smaller ones, a process known as \textit{Ostwald Ripening}. The driving force for Ostwald Ripening is the reduction of surface energy. However, elastic networks hinder the coarsening by imposing an energetic cost on larger droplets. As a result, the droplets achieve an equilibrium size that minimizes the total energy of the quenched mixture. Since the total volume of the system remains unchanged during the quenching and growth of droplets, the energy minimization condition for equilibrium can be expressed as,
\begin{equation} \label{equilibrium1}
    \frac{{\rm d} \Delta \bar{G}(\phi_A, \phi_B,\phi_D,T)}{{\rm d}\phi_B} = 0
\end{equation}
where $\phi_D$ and $\phi_B$ are related by the identity in (\ref{mass_conservation}), and the relationship between $\phi_A$ and $\phi_B$ is given in \eqref{pApB}.

To illustrate the predictions of the model we choose the specific response functions for mixing and elastic free energies. 
For $\Delta \bar{G}_{mix}$, we use the Flory-Huggins theory:
\begin{equation} \label{mixing_energy}
    \Delta \bar{G}_{mix}(\phi_A, \phi_B, \phi_D, T) = (1-\phi_D)\frac{k T}{\nu_m}\bigg\{\frac{\phi_A }{N_A}\ln\phi_A +\phi_B \ln\phi_B+ \chi(T) \phi_B(1-\phi_B)\bigg\} \\
\end{equation}
where  $k$ is the Boltzmann constant, $\nu_m$ is the molecular volume of liquid $B$, $N_A$ is the number of lattice sites occupied by a single chain of liquid $A$ and $\chi(T)$ is the Flory interaction parameter. By inserting the free energy density \eqref{energy1}, along with relations \eqref{Gel},\eqref{Gsur} and \eqref{mixing_energy} into the equilibrium condition (\ref{equilibrium1}), we obtain an  implicit relation for the equilibrium droplet size,
\begin{equation} \label{equilibrium2} 
    \ln\phi_B + (1-\phi_B)(1-1/N_A) +\chi(1-\phi_B)^2 = \frac{\nu_m}{k T}\bigg(\frac{2\Gamma}{r}+ W(r) +\frac{r}{3}W'(r) \bigg)
\end{equation}
where $\phi_B$ is related to the droplet size, $r$,  by the mass conservation constraint \eqref{mass_conservation}, along with the identity \eqref{pD_def}.

Next, we can infer $\chi(T)$ by energy minimization given the knowledge of the saturation volume fraction of liquid $B$ after soaking the matrix in a bath (see Supplementary Information), to write,
\begin{equation}\label{chi}
    \chi = -\frac{\log\phi_{sat} + (1-\phi_{sat})(1-1/N_A)}{(1-\phi_{sat})^2}.
\end{equation}
It is noteworthy that, if the liquid $A$ is a long-chain molecule then $N_A \gg1$, and terms containing $1/N_A$ drop out in equations (\ref{mixing_energy}-\ref{chi}). In the examples discussed subsequently in this paper, we will assume $N_A \gg1$.

For elastic energy calculations, we choose to model the material as  incompressible and use the Mooney-Rivlin constitutive function \cite{mooney1940theory,rivlin1948large} which permits strain-stiffening. The amount of work done in expanding a single droplet from an initial size $r_0$ to final size $r$,  is then given as \cite{raayai2019volume},
\begin{equation} \label{elastic_work}
    W(r) = \frac{4\pi}{3}r^3nE\bigg[\frac{5}{6}-\frac{r_0}{r}- \frac{1}{3}\bigg(\frac{r_0}{r }\bigg)^3+ \frac{1}{2}\bigg(\frac{r_0}{r}\bigg)^4 \bigg]+ \frac{4\pi}{3}r^3(1-n)E\bigg[\frac{r}{2r_0}-\frac{1}{3} -\frac{1}{2}\bigg(\frac{r_0}{r }\bigg)^2  + \frac{5}{6}\bigg(\frac{r_0}{r}\bigg)^3 \bigg]
\end{equation}
where $E$ is the stiffness of the crosslinked polymer and $0 \leq n\leq1$ determines the level of strain stiffening. For $n=1$, the neo-Hookean material model is recovered as the stiffening effect is eliminated. The neo-Hookean limit has been shown to apply only for moderate strains \cite{alan2001engineering}. Additionally, it predicts that the total work done in expanding a cavity arrives at an asymptotic limit as $r\gg r_0$. Hence, for the neo-Hookean model, expansion of all the cavities will only depend on their total volume and not on their radius. To account for nonlinear strain-stiffening effects that emerge at large strains, the Mooney-Rivlin model (\ref{elastic_work}) includes the second term on the right hand side (with $n<1$), which restricts the cavity size. %Namely, a strain-stiffening material model is able to choose an optimum droplet size while the neo-Hookean model is indifferent to the size of the individual droplets. 

An important parameter that appears in equation \eqref{elastic_work} is the initial cavity size, $r_0$, which can be assumed to represent the size of initial defects in the material. For a crosslinked network, it seems natural to choose $r_0$ as the pore size (i.e. $r_0\sim 1$ nm). However, at this scale continuum assumptions break down. Additionally, as
explained by Gent and Tompkins \cite{gent1969surface}, a significant energetic barrier hinders the  expansion of small cavities.  By comparing the contributions of elastic  \eqref{Gel}  and  surface \eqref{Gsur} energies,  it becomes clear that  this barrier is dominated by
surface energy, and appears for $\Gamma/Er_0 \gg 1$.
 In this study, we consider situations in which elastic effects are non-negligible, thus we expect that before elasticity is activated at the continuum scale,  the thermodynamic forces that drive de-mixing, create droplets that have an effective initial size  of $r_0\geq \Gamma/E$, which can be larger than the pore size by orders of magnitude. Note that the precise microscopic mechanisms that drive the formation of the initial pores during the nucleation, should be subject for future work. In this work, we take a representative value of $r_0=0.1\ \mu$m, which obeys the inequality $r_0\geq \Gamma/E$ within the entire range of considered values of $E$ and $\Gamma$. This effective initial size is used to determine the level of circumferential stretch (i.e. $\lambda=r/r_0$) throughout the expansion process at the continuum scale.

%The choice of the initial stress free cavity size $r_0$ is an important model parameter.
%\blue{The surface energy barrier to growth of droplet is inversely proportional to the cavity radius, as can be seen from equation (\ref{equilibrium2}). If the initial cavity size in the crosslinked polymer is too small  ($<10$ nm), then the surface energy barrier not only dominates the elastic energy barrier but also precludes the growth of the droplet. For larger cavity sizes, ($10-1000$ nm) the surface energy barrier and elastic energy barrier may be comparable and the droplet can grow with sufficient quench. In other words, to observe the elastic effects on droplet growth, we require the elastocapillary number associated with the initial cavity size, $\frac{2\Gamma}{Er_0} \leq 1$.}\\

\begin{figure}[h!]
	\centering
  \includegraphics[scale = 0.23]{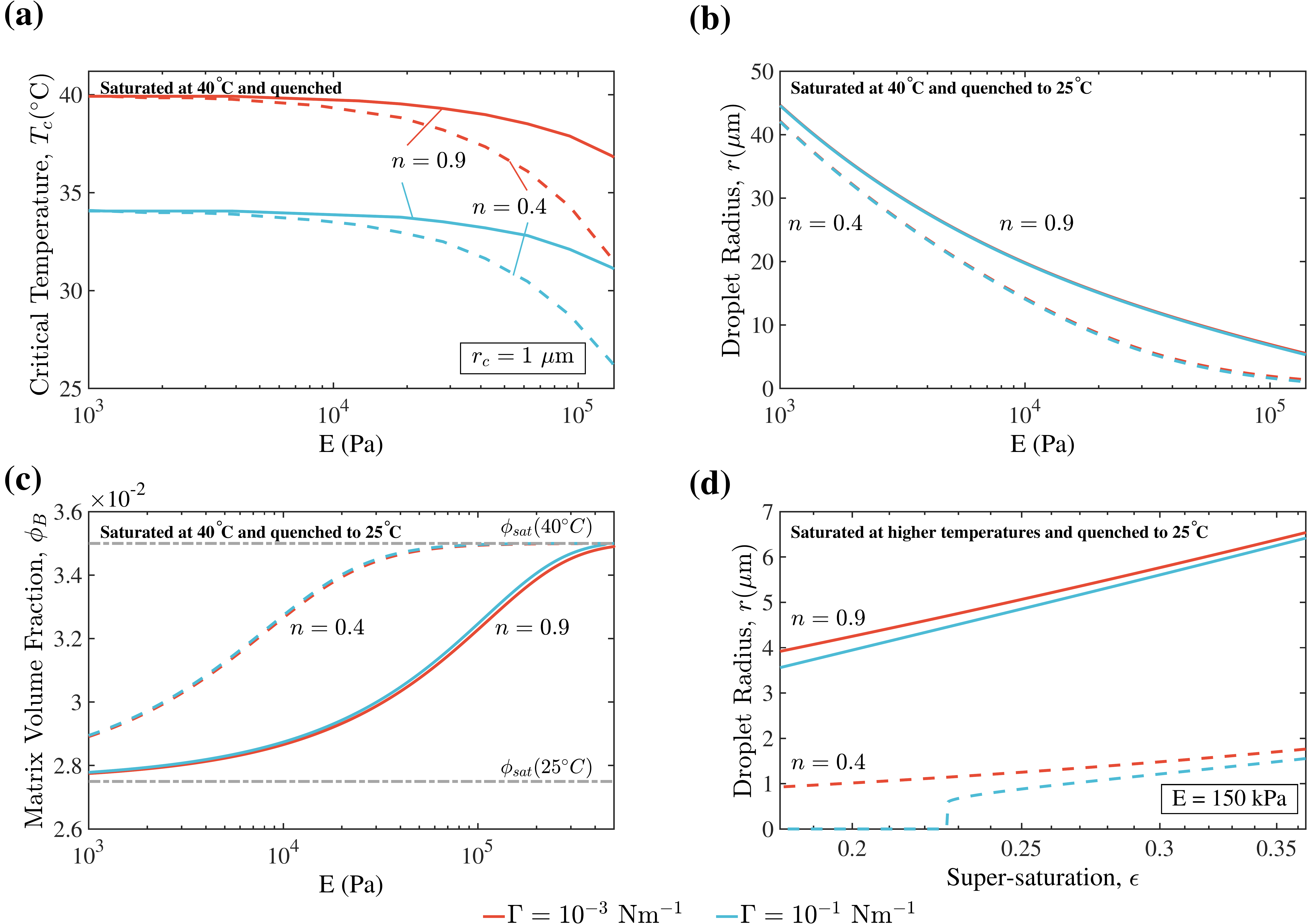}
  \caption{\textbf{Elasticity delays phase separation and arrests Ostwald Ripening.} Model predictions are obtained  for two values of surface tension $\Gamma=10^{-3}$, $10^{-1} \text{N}\text{m}^{-1}$, and two values of the strain stiffening coefficient $n = 0.4$, $0.9$, using the representative model parameters: $\nu_m = 10^{-27}  \ \text{m}^3,\ \phi_{sat} = 0.0005T[^\circ \text{C}] + 0.015,\ n_d = \alpha E, \alpha = 2 \times 10^7 \ \text{N}^{-1} \text{m}^{-1}\text{ and} \ r_0 = 0.1\ \mu \text{m}$.  Quenching occurs from $40^\circ$C to $25^\circ$C unless noted otherwise. (a) The critical  temperature for appearance of droplets of size $r_c = 1\ \mu\text{m}$ in a crosslinked polymer, which is initially saturated at $40^\circ$C, as a function of the elastic modulus.  It is shown that  a deeper quench is required to induce phase separation in a stiffer polymer. Effect of surface tension and strain-stiffening are also noticeable.  (b) Quasi-equilibrium droplet size is shown as a function of the elastic stiffness. A modest effect of strain-stiffening is observed while surface tension has a negligible effect. (c) Volume fraction of liquid $B$ in the matrix is shown as a function of the stiffness and reveals an elastic analogue of Gibbs-Thomson effect; higher stiffness leads to higher solubility. (d) The equilibrium size of droplets scales with the supersaturation.  Here  the system is quenched to $25^\circ$C from different initial temperatures and with $E=150$kPa.  }\label{fig2}
\end{figure}

In the following, we consider example scenarios to illustrate the results of the model. Figure \ref{fig2} shows the short timescale predictions assuming the  matrix is initially saturated with liquid $B$ at  temperature $T_1 = 40^\circ$C. The results are obtained using  representative values of the model parameters $(\nu_m, \phi_{sat},n_d,  \alpha, r_c, r_0)$ to consider the sensitivity of the process to the elastic stiffness  $E$, the surface energy  $\Gamma$, strain stiffening $n$, and the supersaturation, which is defined as $\epsilon = \phi_{sat}(T_1)/\phi_{sat}(T)-1$. Note that within the range of considered temperatures, thermal effects on the  elastic modulus are negligible.

The influence of elasticity and surface energy on the critical temperature for nucleation of droplets with $r_c=1 \ \mu$m is shown in Figure \ref{fig2}(a). As observed, a matrix with higher stiffness and higher strain-stiffening requires a deeper quench to undergo phase separation. Surface energy also increases the barrier to the nucleation of droplets. Similar qualitative behaviors are observed for different values of $r_c$, which can be chosen, for example, as the minimal observable size in an experiment, or the required size in a particular application. Figure \ref{fig2}(b) shows model predictions for the quasi-equilibrium droplet size of a system that is quenched to $T = 25^\circ$C. Higher stiffness of the crosslinked polymer leads to a smaller equilibrium droplet size; a similar trend is followed in the strain-stiffening response whereas  higher strain-stiffening leads to smaller droplet size. Additionally, it is shown that although surface tension plays an important role in determining the critical temperature for appearance of droplets, it does not affect the final droplet size. Figure \ref{fig2}(c) shows the equilibrium concentration of liquid $B$ in the matrix in the quenched state with $25^\circ$C. Analogous to the Gibbs-Thomson effect, elastic stiffness is shown to enhance the solubility of liquid $B$. Two extremes are observed; in the high stiffness limit, characterized by $E\nu_m/ \Delta\chi k T \geq 1$ where $\Delta \chi = \chi(T)-\chi(T_1)$, all of the liquid $B$ that was dissolved at $T_1$ still remains dissolved at the lower temperature $T$, despite the quench. On the other extreme of low stiffness ($E\nu_m/ \Delta\chi k T \ll 1$), elasticity leads to negligible enhancement in the solubility and matrix equilibrium concentration becomes $\phi_{sat}(T)$. 
In Figure \ref{fig2}(d) the matrix is saturated at different initial temperatures $T_1$ and quenched to $25^\circ$C to vary the supersaturation, $\epsilon$. As expected, higher supersaturation leads to larger droplet size, while higher stiffness and higher stiffening result in smaller droplets. If the supersaturation is not high enough (or the quench is not deep enough), the droplets may not appear, consistent with Figure \ref{fig2}(a).

%%%%%%%%%%%%%%%%%%%%%%%%%%%%%%%%%%%%%%%%%%%%%%%%%%%%%%%%%%%%%%%%%%%%%%%%%%%%%%%%

%%%%%%%%%%%%%%%%%%%%%%%%%%%%%%%%%%%%%%%%%%%%%%%%%%%%%%%%%%%%%%%%%%%%%%%%%%%%%%%

\subsection*{\textbf{Long Timescale: Propagation of dissolution front in graded elastic polymers} }\label{section_2.2}
\begin{figure}[h]
	\centering
  \includegraphics[scale = 0.5]{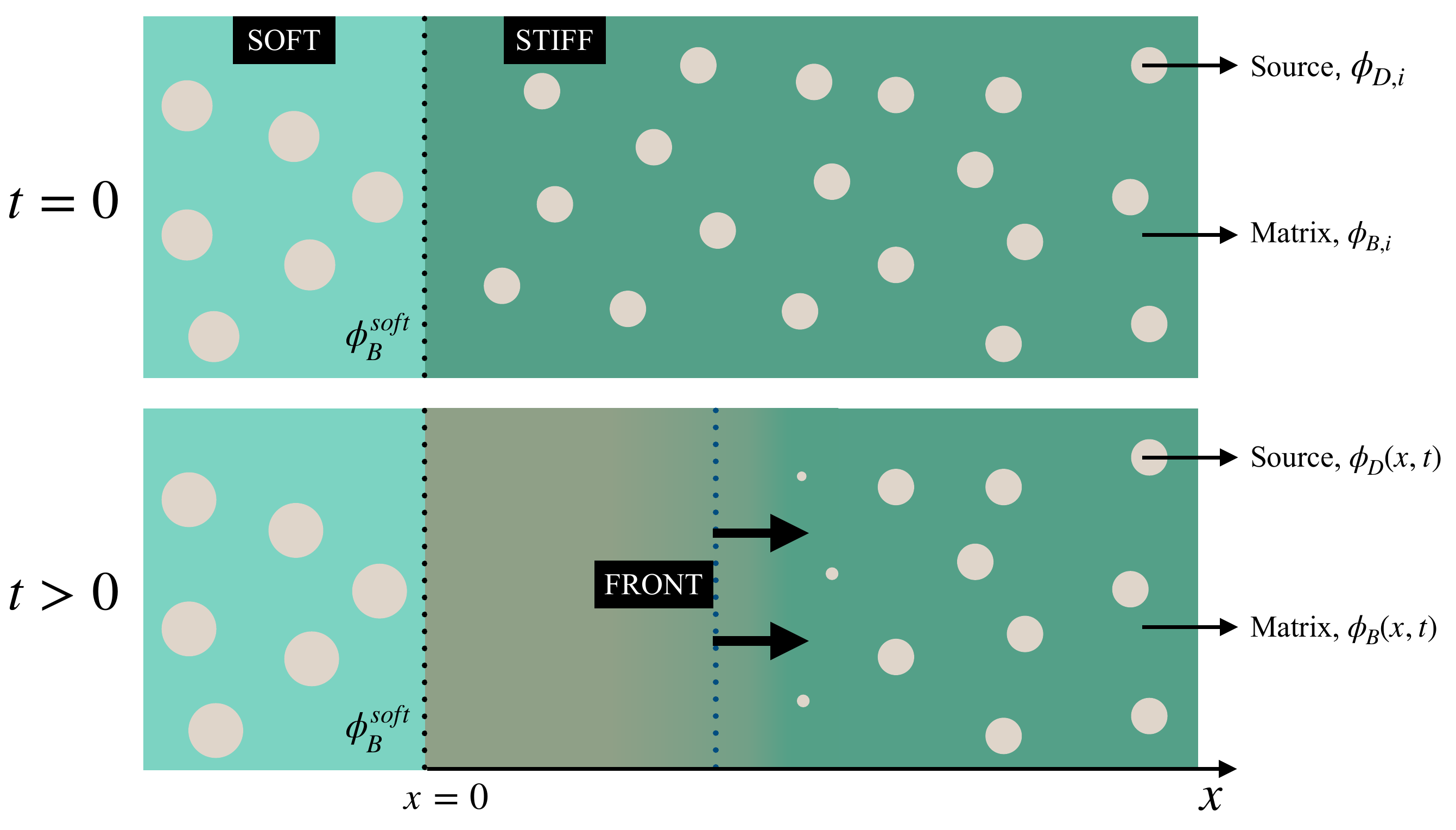}
  \caption{\textbf{Schematic of the Front Propagation Process.} The graded polymer is made of soft and stiff sides that span indefinitely along the negative and positive directions of the longitudinal coordinate $x$, respectively. At short timescale ($t=0$), the system is in the quenched quasi-equilibrium state with droplets distributed throughout. At the long timescale ($t>0$), the system exhibits long-range migration of liquid $B$ and a resulting closure of droplets.}\label{fig3}
\end{figure}

In this section, we build on the quasi-equilibrium solution obtained at the short timescale to understand the long-term dynamics of phase separation in spatially heterogeneous materials. Inspired by the experimental work in \cite{rosowski2020elastic}, we consider the simplest case of spatial heterogeneity - a uniformly soft and a uniformly stiff polymer joined together (Fig. \ref{fig3}). We assume that quenching is rapid enough so that the short- and long-timescale processes are well separated. As a consequence, the nucleation and growth of droplets to their quasi-equilibrium size is not influenced by the composite nature of the polymer. The droplet distribution on either side is the same as it were in the uniform case (without composite sample), and depends only on the local properties. The short-timescale solution thus serves as the initial condition for the dynamic study. As established earlier, elasticity enhances the solubility, therefore, the stiffer side has higher concentration of liquid $B$ in the matrix as compared to the softer side. This concentration difference between the two sides sets up a diffusive flux of liquid $B$ from higher to lower concentration. As the liquid in the matrix depletes on the stiff side, the droplets serve as a source of liquid $B$ to replenish the matrix. In the analysis, we focus only on the dynamics of the stiff side since that is where we expect the dissolving droplets to create a propagating front. The effect of the soft side is accounted for through the boundary conditions.  \\

We construct a 1-D picture of the above process. The spatial variable is denoted by $x$; the soft side, of stiffness $E_{soft}$, spans $x\in(-\infty,0)$ and the stiff side, of stiffness $E_{stiff}$, spans $x\in(0,\infty)$. The matrix and droplet concentrations are now functions of space and time (i.e. $\phi_B=\phi_B(x,t)$ and $\phi_D=\phi_D(x,t)$). Due to the long-range migration of liquid $B$, the mass conservation constraint (\ref{mass_conservation}) no longer holds and $\phi_B(x,t)$ and $\phi_D(x,t)$ are independent variables. We model the droplets as an \textit{exhaustible}, spatially distributed source in the diffusion equation. The dynamical system can be formulated as,
\begin{align}
    \pards{\phi_B}{t} &= D\frac{\partial^2 \phi_B}{\partial x^2}+ s(\phi_B,\phi_D)  \label{dyneqn1}\\
\pards{\phi_D}{t} &= -s(\phi_B,\phi_D)  \label{dyneqn2}
\end{align}
 where $D$ is the diffusion constant and $s(\phi_B,\phi_D)$ is the source term. The initial conditions are supplied from the short timescale solution,
\begin{equation}\label{ic}
    \phi_B(x>0,0) = \phi_{B,i} \quad \text{and} \quad \phi_D(x,0) = \phi_{D,i}
\end{equation}
where $\phi_{B,i}$ is the matrix volume fraction of liquid $B$ at the short-timescale which is in equilibrium with the volume fraction $\phi_{D,i}$ of the droplets. All the liquid $B$ that leaves the stiff side, permeates the soft side which requires continuity of flux at the $x=0$ interface; additionally we require that the flux decays at the remote end ($x \rightarrow \infty$), to write the boundary conditions, 
\begin{align}\label{bc}
    \pards{\phi_B}{x}(0^+,t) &= D(\phi_B(0^+,t)-\phi_B^{soft})\\ \pards{\phi_B}{x}(\infty,t) &= 0
\end{align}
where $\phi_{B}^{soft}$ is the volume fraction of liquid $B$ in the soft side at the short timescale. Experiments \cite{rosowski2020elastic} have indicated that  liquid $B$ that migrates from the stiff side to the soft side, is rapidly taken up by the droplets on the soft side, therefore we assume that $\phi_{B}^{soft}$ remains constant over the long timescale. 

The key feature in the above model is the kinetic relation \eqref{dyneqn2}, which determine the  flow of liquid $B$ from the droplets to the matrix. At the short timescale, the thermodynamic force for phase separation is balanced by the resistance due to elastic and surface energies. When this equilibrium is disturbed by the migration of liquid $B$ in the matrix, the thermodynamic force for phase separation becomes weaker and the droplets start to mix back. This process is controlled by the  source term $s(\phi_B,\phi_D)$, which is derived next. 

Considering a unit element of the phase separated material system, the local process of expelling fluid from the droplets at rate $\partial V_D/\partial t$, results in the increase of liquid $B$ in the matrix at the rate $\partial{ V}_B/\partial t=-\partial{ V}_D/\partial t$. The corresponding change in free energy must obey the second law of thermodynamics, and thus results in the inequality
\begin{equation}\label{2ndLaw}
    \frac{\partial{\Delta G}}{\partial t}=\left(\frac{\partial \Delta G}{\partial V_D}-\frac{\partial \Delta G}{\partial V_B}\right)\frac{\partial{V}_D}{\partial t}\leq 0
\end{equation}
which holds for  constant values of $V_A$ and $T$.
Accordingly, we define the \textit{driving force for droplet dissolution}, readily written in terms of volume fractions, as 
\begin{equation}\label{df}
    f(\phi_A,\phi_B)=\frac{\partial \Delta G}{\partial V_D}-\frac{\partial \Delta G}{\partial V_B}=\frac{\partial \Delta \bar{G}}{\partial \phi_D}-\left(\frac{1-\phi_B}{1-\phi_D}\right)\frac{\partial \Delta \bar{G}}{\partial \phi_B}
\end{equation}
This driving force (per unit volume of the material system) serves as the thermodynamic conjugate to the rate of dissolution ${\partial V}_D/\partial t$, or equivalently, to the rate of change of droplet volume fraction $\partial \phi_D/\partial t$.

In light of  the inequality \eqref{2ndLaw},  the kinetic relation \eqref{dyneqn2} that governs the rate of dissolution can be chosen as   $\partial\phi_D/\partial t =-\mathcal{S}(f)$, where the source  function  $\mathcal{S}(f)$ obeys the inequality $f\mathcal{S}(f)\geq0$, for all values of $f$. While $\mathcal{S}(f)$ can take an arbitrary form, the results of this work are obtained using a linear relation $\mathcal{S}(f)=Kf$ with the constant kinetic parameter $K>0$, which applies for small departures from thermodynamic equilibrium. By inserting \eqref{energy1}, in \eqref{df}, this source term further specializes to the form
\begin{equation} \label{source}
	s(\phi_B, \phi_D) = K\bigg\{\frac{2\Gamma}{r}+ W(r) +\frac{1}{3}W'(r) -\frac{kT}{\nu_m}\bigg(\ln\phi_B + (1-\phi_B)(1-1/N_A) +\chi(1-\phi_B)^2 \bigg)\bigg\} H(\phi_D)
\end{equation}
where $s(\phi_B,\phi_D)={\mathcal{S}}(f(\phi_B,\phi_D))$
and the Heaviside function $H(\phi_D)$ ensures the source is exhausted when $\phi_D = 0$. Note that, as expected, in the initial state the driving force vanishes, as seen by comparing the above relation with the equilibrium condition \eqref{equilibrium2}.

{To determine the effect of the kinetic parameter, $K$, on the dynamics, we return to examine the governing equations \eqref{dyneqn1} and \eqref{dyneqn2}. First, it is straight forward to determine a dimensionless counterpart to $K$, as
\begin{equation}
    \mathcal{K}=\frac{kTL^2K}{D\nu_m} %=\tc{}{\frac{EL^2K}{D}}
\end{equation}
where $L$ is a characteristic length scale of the system. We refer to $\mathcal{K}$ as the \textit{dissolution number}. As observed by comparing the two terms on the right hand side of  \eqref{dyneqn1}, if $\mathcal{K} \gg1$ the source is rapidly exhausted and the propagation of the front relies on the diffusion.  Hence, we refer to the front propagation at this  limit as \textit{diffusion limited}. On the other hand, if $\mathcal{K} \ll1$, the release of liquid lingers while  liquid $B$ migrates at a faster rate.} Hence, we refer to front propagation at this limit as \textit{dissolution limited}.

\begin{figure}[H]
	\centering
  \includegraphics[scale = 0.4]{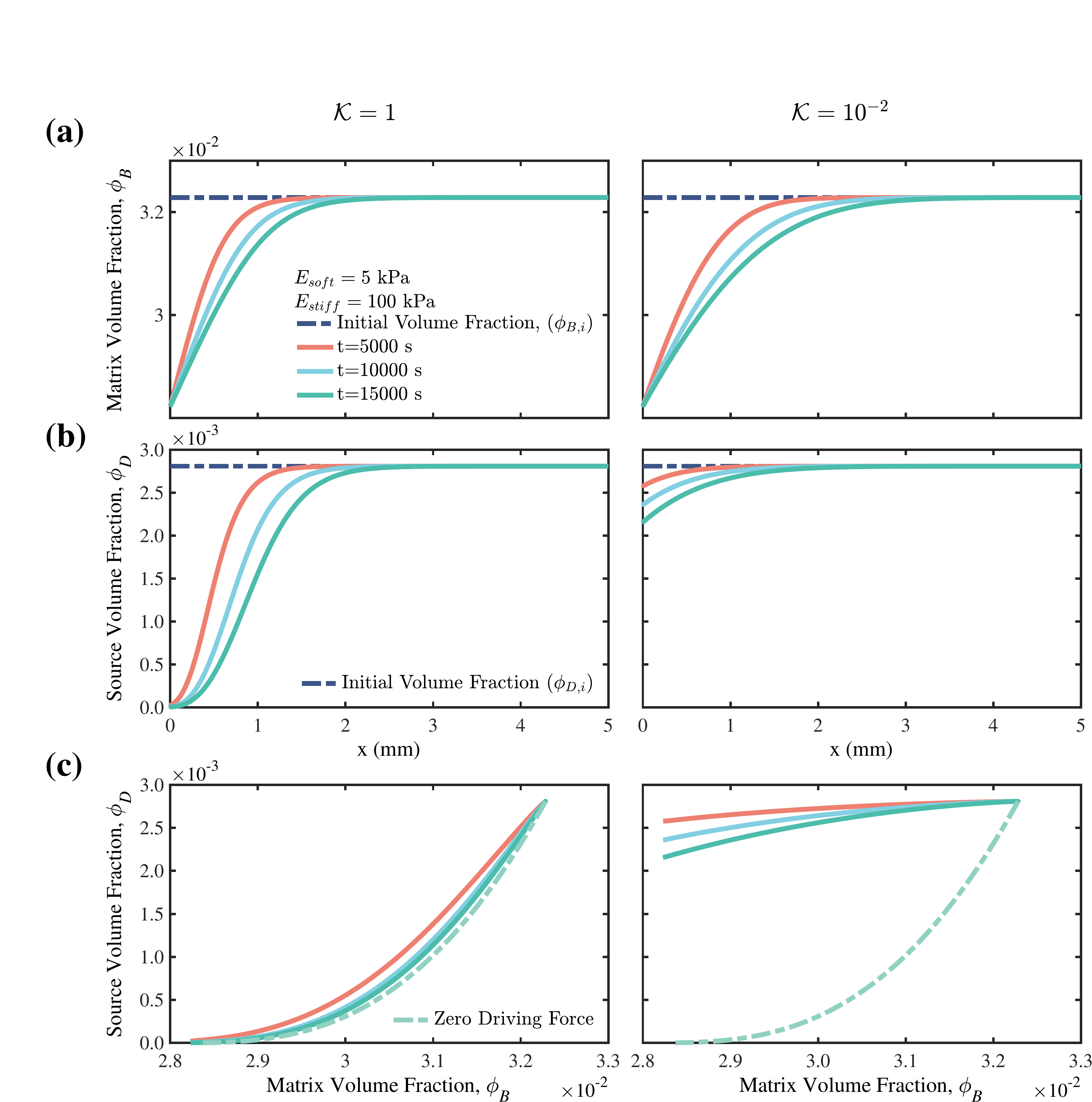}
  \caption{\textbf{Front Propagation.} Predictions are shown for long range migration of liquid $B$ from a stiff polymer $(E_{stiff} = 100$ kPa) to a soft polymer ($E_{soft} = 5$ kPa), both with strain-stiffening coefficient $n = 0.9$. The model parameters are: $D = 5\times 10^{-11} \ \text{m$^2$}\text{s}^{-1}$, $L = 10^{-3} \ \text{m},\  \Gamma=10^{-3} \ \text{N}\text{m}^{-1}, \ \nu_m = 10^{-27}  \ \text{m}^3,\ \phi_{sat} = 0.0005T[^\circ \text{C}] + 0.015,\ n_d = \alpha E, \alpha = 2 \times 10^7 \ \text{N}^{-1} \text{m}^{-1} \text{ and } \ r_0 = 0.1\ \mu \text{m}  $. Both sides were saturated at $40^\circ$C and then quenched to $25^\circ$C, which decides their initial conditions. Results are shown for $\mathcal{K} = 1, 10^{-2}.$ (a) Volume fraction of liquid $B$ in the matrix is shown for the stiff side, with the interface at $x=0$. The liquid $B$ migrates from stiff to soft side. (b) The source volume fraction, representative of the size of the droplets, visually represents a dissolution front progressing further into the stiff side over time. (c) The system approaches zero driving force at different rates based on $\mathcal{K}$.}\label{fig4}
\end{figure}

Now, to illustrate the effect of elasticity on the long-timescale behavior, we consider an example scenario. The time-evolution of a system following quenching from $40^{\circ}$C to $25^{\circ}$C is portrayed in Figure \ref{fig4} by the volume fractions on the stiff side at various times. Curves are obtained using a representative set of model parameters $(D, \mathcal{K}, E_{soft}, E_{stiff}, \Gamma, \nu_m, \phi_{sat},n_d,  \alpha, r_0)$. Figure \ref{fig4}(a) shows that the matrix volume fraction displays a characteristic diffusive profile, governed by the smaller volume fraction on the soft side. The source (Fig. \ref{fig4}(b)) is activated to replenish the matrix until it is depleted, and the front appears to propagate into the stiff side (i.e along the positive $x$ direction). For smaller values of $\mathcal{K}$, source is slower in giving up liquid $B$ resulting in dissolution-limited front propagation. Figure \ref{fig4}(c) illustrates the origin of $\mathcal{K}$-based differences by looking at the boundary where the driving force vanishes in the $(\phi_B, \phi_D)$ plane. This curve represents the locus of quasi-equilibrium states given by $s(\phi_B,\phi_D) = 0$, for a fixed stiffness. For $\mathcal{K}=1$, the system quickly reaches the zero driving force, implying that source continually gives up more liquid to maintain equilibrium with the depleting matrix. In contrast, for $\mathcal{K} = 10^{-2}$, the source response is much slower and is not able to achieve a rapid equilibrium with the matrix, thus making the dissolution of the droplets as the process that controls the front propagation.

% \begin{figure}[H]
% 	\centering
%   \includegraphics[scale = 0.35]{matrixandsource_concentration.pdf}
%   \caption{\textbf{Front Propagation.} Predictions are shown for long range migration of liquid $B$ from a stiff polymer $(E_{stiff} = 100$ kPa) to a soft polymer ($E_{soft} = 5$ kPa), both with strain-stiffening coefficient $n = 0.9$. The model parameters are: $D = 5\times 10^{-11} \ \text{m$^2$}\text{s}^{-1}$ and $K = 5\times 10^{-10} \ \text{Pa}^{-1}\text{s}^{-1}, \ \nu_m = 10^{-28}  \ \text{m}^3,\ \phi_{sat} = 0.0005T[^\circ \text{C}] + 0.015,\ n_d = \alpha E,\ r_0 = 0.1\ \mu \text{m} \text{ and }  \alpha = 2 \times 10^7 \ \text{N}^{-1} \text{m}^{-1}$. Both sides were saturated at $40^\circ$C and then quenched to $25^\circ$C, which decides their initial conditions. (a) Volume fraction of liquid $B$ in the matrix is shown for the stiff side, with the interface at $x=0$. The liquid $B$ migrates from stiff to soft side. (b) The source volume fraction, representative of the size of the droplets, visually represents a mixing front progressing deeper into the stiff side over time.}\label{fig4}
% \end{figure}%

\begin{figure}[H]  %
    \centering
    \includegraphics[scale = 0.3]{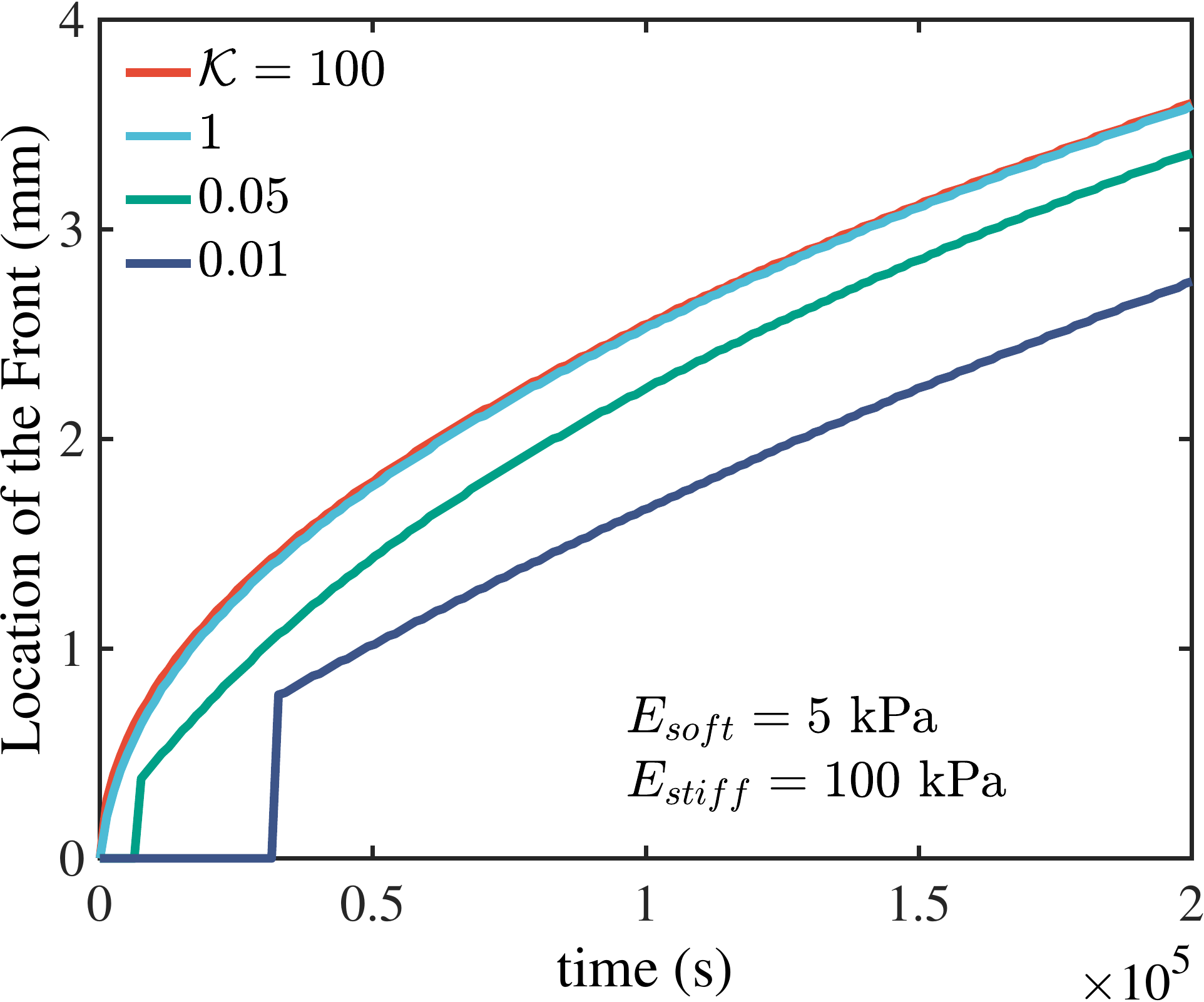} %
    \caption{\textbf{Front Location.} Model predictions are shown 
    for the location of the front. The model parameters are: $D = 5\times 10^{-11} \ \text{m$^2$}\text{s}^{-1}$, $\Gamma=10^{-3} \ \text{N}\text{m}^{-1}, \ \nu_m = 10^{-27}  \ \text{m}^3,\ \phi_{sat} = 0.0005T[^\circ \text{C}] + 0.015,\ n_d = \alpha E,\ \alpha = 2 \times 10^7 \ \text{N}^{-1} \text{m}^{-1} \text{ and } \ r_0 = 0.1\ \mu \text{m}$. Both sides were saturated at $40^\circ$C and then quenched to $25^\circ$C, which decides their initial conditions.}%
    \label{figx}%
\end{figure}

It is now possible to examine the effect of model parameters on the propagation of the front. We define the \textit{location} of the front as the $x$ coordinate where the source volume fraction is half of its initial value. The front starts propagating from $x=0$ in the positive $x$ direction; its location for different values of $\mathcal{K}$ is shown in Figure \ref{figx}. The location of the front follows $x = At^{1/2}$ for $\mathcal{K} \geq 1$. This behavior is typical of diffusive fronts and thus, the front location is expected to follow this in the diffusion limited propagation. For $\mathcal{K} \ll 1$, the location trend approaches the $t^{1/2}$ limit in long-time behavior, but shows initial deviations.

\begin{figure}[H]  %
    \centering
    \includegraphics[scale = 0.3]{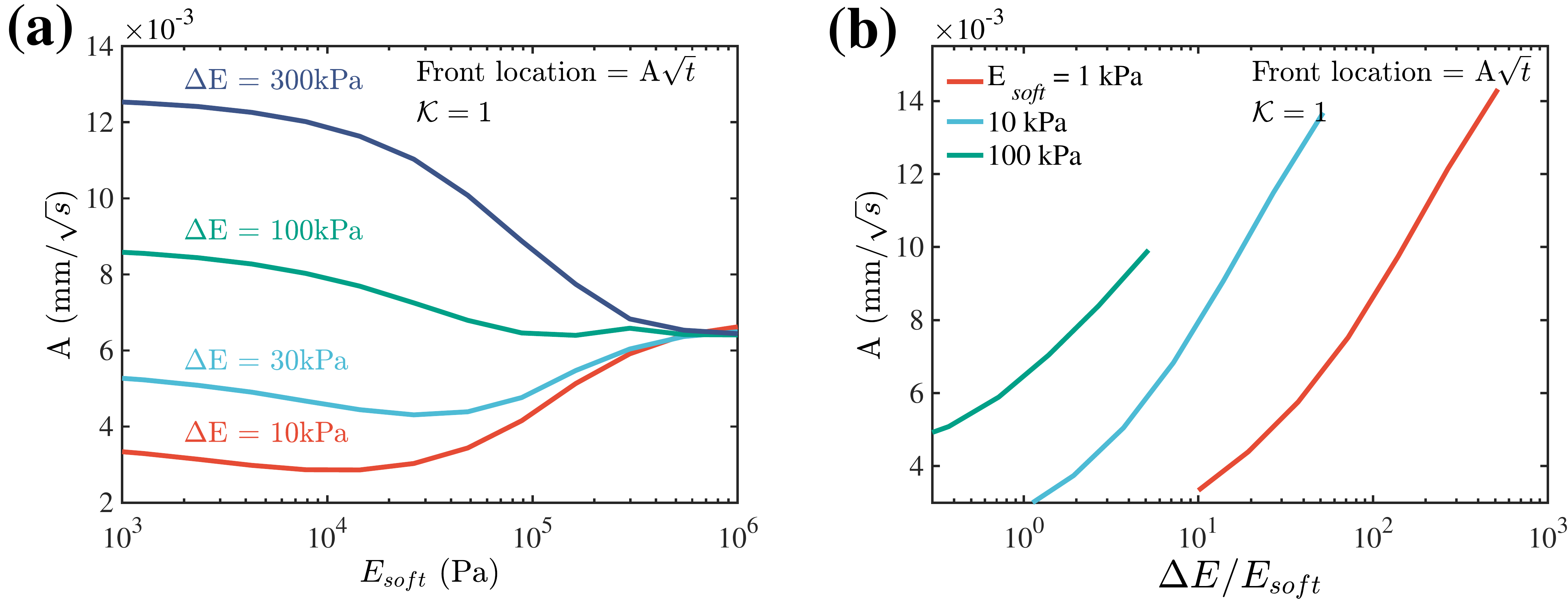} %
    \caption{\textbf{Elastic Properties Have a Nonlinear Effect on Front Speed.} Model predictions are shown 
    for the parameter $A$ that determines the front speed. The model parameters are: $D = 5\times 10^{-11} \ \text{m$^2$}\text{s}^{-1}$, $\Gamma=10^{-3} \ \text{N}\text{m}^{-1}, \ \nu_m = 10^{-27}  \ \text{m}^3,\ \phi_{sat} = 0.0005T[^\circ \text{C}] + 0.015,\ n_d = \alpha E, \ \alpha = 2 \times 10^7 \ \text{N}^{-1} \text{m}^{-1} \text{ and } \ r_0 = 0.1\ \mu \text{m}$. Both sides were saturated at $40^\circ$C and then quenched to $25^\circ$C, which decides their initial conditions. The speed of the 50\%-depletion front (a) changes non-monotonically with stiffness of the soft-side, while keeping $\Delta E$ fixed, (b) increases monotonically with $\Delta E$, while keeping the soft-side stiffness constant.}%
    \label{fig5}%
\end{figure}

The choice of stiffnesses is an important factor that governs the speed, or equivalently, the coefficient $A$. Results for the example case with representative values of model parameters $(D, \mathcal{K}, E_{soft}, \\ E_{stiff}, \Gamma, \nu_m, \phi_{sat},n_d,  \alpha, r_0)$ are shown in Figure \ref{fig5}. Increasing the stiffness of both the soft and the stiff sides while keeping fixed their difference ($\Delta E = E_{stiff} -E_{soft} $), changes the speed non-monotonically as seen in Figure \ref{fig5}(a). This challenges the current understanding that the front speed only depends on the stiffness difference. The front speed, in fact, also depends on the individual stiffness of either side and this may lead to a scenario where the front speeds approach the same value regardless of the $\Delta E$. Figure \ref{fig5}(b) shows that if $E_{soft}$ is held constant, the speed increases nonlinearly with increasing $\Delta E$.

\subsection*{\textbf{Experimental Validation} }\label{section_2.4}
\begin{figure}[H]
\centering
  \includegraphics[scale = 0.23]{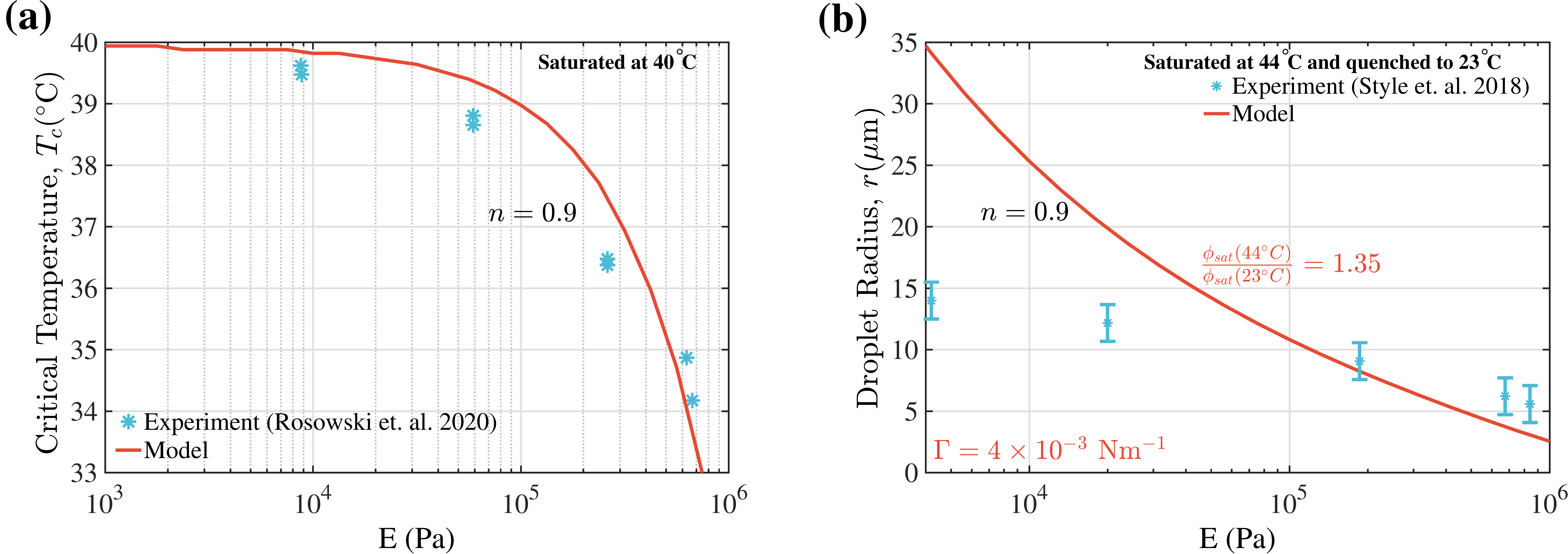}
  \caption{\textbf{Comparison of Model Predictions with Experiments} Model predictions are compared with experimental results from \cite{rosowski2020elastic} with the following model parameters: $n = 0.9, \ \nu_m = 3.7\times 10^{-28}  \ \text{m}^3,\ N_A \gg 1,\ \phi_{sat} = 0.000467T[^\circ C] + 0.01728, \Gamma = 4\times 10^{-3} \ \text{N}\text{m}^{-1}, \ n_d = \alpha E \text{ and } \alpha = \  1.4\times 10^{7} \ \text{N}^{-1}\text{m}^{-1}$. The saturation temperatures are indicated on the figures. (a) Critical temperature prediction shows good agreement with experiments for a moderate strain-stiffening elastic response (b) Model predictions show good agreement for high stiffness and overprediction for smaller stiffness.  }\label{fig6}
\end{figure}

We carry out the validation of our model by comparison with the experimental results reported in \cite{style2018liquid}. We use a set of parameters reported in the study ($\nu_m,\ N_A,\ \phi_{sat}, \ \Gamma, \ n_d$) along with $n=0.9$, $r_c = 1 \ \mu$m and $r_0 = 0.1\ \mu $m for the Mooney-Rivlin constitutive model for the matrix. Figure \ref{fig6}(a) shows the prediction for critical temperature, $T_c$, for different stiffnesses and is in agreement with the experimental results. The predictions for equilibrium droplet size are in good agreement with the experimental observations in the high-stiffness range (Fig. \ref{fig6}(b)). For softer samples, experiments reported polydisperse droplet size and we speculate this to be the reason for the observed discrepancy in the low-stiffness regime. For the front propagation at the longer timescale, while there is not enough experimental data to quantitatively compare with model predictions, we find qualitative agreement between the limited experimental data and model predictions for $\mathcal{K}=1$. Further experimental investigation of  front speeds with varying elastic properties is necessary to fully understand the nonlinear elastic effects and the various regimes of front propagation.

\section*{Discussion}
% \blue{This section needs to relate/discuss the significance of the above work in the context of biological systems, other polymer systems to make it a broad appeal. Can also discuss aspects of controlling the droplet pattern by designing the material stiffness}
% \color{black}
 In this work, we study the effect of elasticity on the phase separation of binary liquid mixtures and the resulting long-range species migration in heterogeneous materials. A thermodynamically-consistent theoretical model that captures the influence of nonlinear elastic properties on both the formation  of precipitates and the subsequent process of their dissolution across a propagating front, is proposed. To our knowledge, no systematic theoretical study of elastic effects on the formation and dynamics of \textit{liquid precipitates} in soft nonlinear elastic solids has been carried out till date.

% In absence of elasticity, the phase-separated mixtures exhibit coarsening until both phases are completely separated. However, elasticity changes the energy landscape to arrest the Ostwald Ripening by creating a stable equilibrium state of dispersed liquid droplets throughout the matrix. Further, elasticity can significantly delay the phase-separation. A highly nonlinear, stiff material can show significant decrease in the nucleation temperature. 

In typical liquid-liquid phase separation, once the liquid droplets cross the energy barrier imposed by surface energy, they can continue to grow. However, in the presence of an elastic matrix, the droplet has to work against the elastic resisting force to activate large  deformations, which increases both the nucleation barrier as well as the energetic cost of growing the droplet. Our model reveals the  contribution of both surface tension and elastic properties to the nucleation barrier and the quasi-equilibrium separated state. The equilibrium droplet size is determined by minimization of free energy and is shown to scale with the supersaturation. Stiffer and strain-stiffening materials show smaller droplet sizes due to their higher solubility of the minority component. We show that these effects follow nonlinear trends. The analysis assumes that the short-timescale process of phase separation is well separated from the long-timescale process of species migration; this is ensured by requiring a rapid quench. 

The long-time dynamics that can arise in presence of material heterogeneity are studied through an illustrative example of a graded sample comprising a stiff and a soft polymer joined together. For this example, diffusion of the minority component is affected by a concentration mismatch, and leads to formation of a dissolution front. A thermodynamic argument is employed to model the chemo-mechanically coupled kinetics of dissolution, which are found to be governed by the dimensionless dissolution number $(\mathcal{K})$. Distinct regimes of \textit{diffusion dominated} and \textit{dissolution dominated} front propagation are shown to emerge, for different ranges of $\mathcal{K}$.   Elasticity is shown to govern the speed of the front by controlling the quasi-equilibriated matrix volume fraction of the minority component on either side of the graded polymer. If the droplet volume is large, the source (i.e. droplet content) is large, thereby taking longer to deplete it and vice versa. Accordingly, the front moves faster in stiffer materials where the higher solubility implies smaller droplet volume. Since the volume fraction of the minority component depends nonlinearly on the stiffness, the speed of the front also shows a nonlinear trend. In fact, it not only exhibits a non-monotonic dependence on the stiffness difference between the soft and stiff sides $(\Delta E)$, but is  also shown to depend separately on the individual stiffnesses. 

%Keeping the soft side at a  fixed stiffness, the speed is shown to monotonically increase with $\Delta E$. However, keeping $\Delta E$ fixed, the speed initially decreases, attains a minimum and then increases with the soft side stiffness.

While in this paper our focus is on the effect of elasticity on the process of phase separation, a number of other variables can play similar role in controlling the solubility and migration of species. For instance, local changes in temperature can affect the solubility and the Flory interaction parameter $(\chi)$, which affects the droplet size at the short timescale and, in turn, dictates the long timescale migration. In this way it is possible to reverse the Ostwald Ripening process in order to grow smaller droplets at the cost of larger ones, as observed recently \cite{rosowski2020elastic2}. Dissimilar strain-stiffening behavior can also significantly affect the stresses in the matrix, which directly affects both the short- and long-timescale behavior. As an example, a material with higher strain-stiffening will show smaller droplet size when compared to another material with the same stiffness, leading to differences in long-time behavior. Application of external stresses on the system can also have a similar effect.  The model presented here can provide guidance to quantitatively determine the appropriate combination of these factors to achieve a desired outcome.

% \noindent Through this work we have demonstrated the role of elasticity in controlling the phase separation process. While we have focussed on thermally induced phase separation, the same theory can be generalized to include other means of phase separation like external stresses. Lastly, we consider small volume fractions of the dilute phase in the mixture which allows us to discount elastic interaction between the droplets. However, for softer materials where the model predicts large droplet sizes, neglecting elastic interaction may not be accurate.\\

Finally, this work highlights the importance of elasticity as a potential regulating mechanism for controlling solubility and phase-separation in various physical settings. Tuning the elastic properties of the matrix material can impart control over the stable droplet distribution for industrial applications, it can  drive the formation of membraneless organelles in biological settings, and can affect the transport of potent greenhouse gases in soil. Although  these systems may differ in lengthscale and diffusion properties, given the elastic properties, their long-time evolution  can be qualitatively predicted based on the value of their  dissolution number.  Future work should extend to more complex heterogeneous settings, and should provide additional insights into the early stages of droplet nucleation and growth, before continuum scale is attained.

\section*{Acknowledgements}
The authors would like to thank Eric R. Dufresne and Robert W. Style for introducing us to this problem. The authors would like to acknowledge helpful discussions with Prashant K. Purohit and Vaishali Garga.

\section*{Author Contributions} 
M. Kothari and T. Cohen conceived the mathematical models,
interpreted the computational results, and wrote the paper. M. Kothari performed the numerical simulations.
All authors gave final approval for publication.

\section*{Competing Interests} Inapplicable to this manuscript.

%%%%%%%%%%%%%%%%%%%%%%%%%%%%%%%%%%%%%%%%%
\newpage 

\section*{Supplementary Information} \label{Methods}%
We follow the same notation as in the main text. 
\subsection*{\textbf{Flory Interaction Parameter}}
Upon soaking the polymer in a bath of liquid $B$, it is saturated and reaches equilibrium. Since the saturated state is a homogeneously mixed state, $V_D = 0$ and $V =V_M$. The change in free energy upon soaking the polymer in liquid $B$ is given as,
\begin{align} 
    \Delta G_{mix} = &V\frac{k T}{\nu_m}\bigg\{\frac{\phi_A }{N_A}\ln\phi_A +\phi_B \ln\phi_B+ \chi(T) \phi_B(1-\phi_B)\bigg\}\\
    =&\frac{k T}{\nu_m}\bigg\{\frac{V_A }{N_A}\ln\frac{V_A}{V_0+V_B} +V_B \ln\frac{V_B}{V_0+V_B}+ \chi(T) V_B\bigg(1-\frac{V_B}{V_0+V_B}\bigg)\bigg\}
    % =&\frac{V_0}{1-\phi_B}\frac{k_B T}{\nu_m}\bigg\{\frac{\phi_A }{N_A}\ln\phi_A +\phi_B \ln\phi_B+ \chi(T) \phi_B(1-\phi_B)\bigg\} \\
    % \Delta \bar{G}_{mix}= \frac{\Delta G_{mix}(\phi_A, \phi_B,T)}{V_0} = &\frac{1}{1-\phi_B}\frac{k_B T}{\nu_m}\bigg\{\frac{\phi_A }{N_A}\ln\phi_A +\phi_B \ln\phi_B+ \chi(T) \phi_B(1-\phi_B)\bigg\}
\end{align}
The saturation volume fraction $\phi_{sat}$ of liquid $B$ is obtained by minimizing $\Delta G_{mix}$ with $V_B$, 
\begin{equation}
    \frac{\partial \Delta {G}_{mix}}{\partial V_B}\bigg\vert_{\phi_B = \phi_{sat}} =0  
\end{equation}
which gives,
\begin{equation}
    \chi = -\frac{\log\phi_{sat} + (1-\phi_{sat})(1-1/N_A)}{(1-\phi_{sat})^2}
\end{equation}

\subsection*{\textbf{Equilibrium Droplet Size}}
We consider the crosslinked polymer saturated with liquid $B$ to a volume fraction $\phi_{sat}(T_1)$ at a temperature $T_1$. The assembly is then quenched to a lower temperature $T$. Since we consider a closed system during quenching, the total volume of liquid $B$ is the same as when the system was saturated, i.e. $V_B+V_D = \phi_{sat}(T_1)V$. Now, the total free energy change upon phase separation can be given as,
\begin{multline}
    \Delta {G} = \frac{k T}{\nu_m}\bigg\{\frac{V_A }{N_A}\ln\frac{V_A}{V_0+V_B} +V_B \ln\frac{V_B}{V_0+V_B}+ \chi(T) V_B\bigg(1-\frac{V_B}{V_0+V_B}\bigg)\bigg\} + V_D\bigg\{W(r)+\frac{3\Gamma}{r}\bigg\}.%
\end{multline}
The equilibrium equation, 
\begin{equation}
     \frac{\rm d \Delta{G}}{\rm d V_B}  =0,
\end{equation}
gives the following implicit relation for the equilibrium droplet size,
\begin{equation} 
    \ln\phi_B + (1-\phi_B)(1-1/N_A) +\chi(1-\phi_B)^2 = \frac{\nu_m}{k T}\bigg(\frac{2\Gamma}{r}+ W(r) +\frac{r}{3}W'(r) \bigg),
\end{equation}
where $\phi_B$ is related to the droplet size, $r$,  by the mass conservation constraint (equation (6) in the maintext).
%%%%%%%%%%%%%%%%%%%%%%%%%%%%%%%%%%%%%%%%%


\begin{thebibliography}{10}
\expandafter\ifx\csname url\endcsname\relax
  \def\url#1{\texttt{#1}}\fi
\expandafter\ifx\csname urlprefix\endcsname\relax\def\urlprefix{URL }\fi
\expandafter\ifx\csname href\endcsname\relax
  \def\href#1#2{#2} \def\path#1{#1}\fi

\bibitem{brangwynne2009germline}
C.~P. Brangwynne, C.~R. Eckmann, D.~S. Courson, A.~Rybarska, C.~Hoege,
  J.~Gharakhani, F.~J{\"u}licher, A.~A. Hyman, Germline p granules are liquid
  droplets that localize by controlled dissolution/condensation, Science
  324~(5935) (2009) 1729--1732.

\bibitem{kothari2019thermo}
M.~Kothari, S.~Niu, V.~Srivastava, A thermo-mechanically coupled finite strain
  model for phase-transitioning austenitic steels in ambient to cryogenic
  temperature range, Journal of the Mechanics and Physics of Solids 133 (2019)
  103729.

\bibitem{tancret2018phase}
F.~Tancret, J.~Laigo, F.~Christien, R.~Le~Gall, J.~Furtado, Phase
  transformations in fe--ni--cr heat-resistant alloys for reformer tube
  applications, Materials Science and Technology 34~(11) (2018) 1333--1343.

\bibitem{smith2016phase}
T.~Smith, B.~Esser, N.~Antolin, A.~Carlsson, R.~Williams, A.~Wessman,
  T.~Hanlon, H.~Fraser, W.~Windl, D.~McComb, et~al., Phase transformation
  strengthening of high-temperature superalloys, Nature communications 7~(1)
  (2016) 1--7.

\bibitem{nabarro1940influence}
F.~Nabarro, The influence of elastic strain on the shape of particles
  segregating in an alloy, Proceedings of the Physical Society 52~(1) (1940)
  90.

\bibitem{doi1985effects}
M.~Doi, T.~Miyazaki, T.~Wakatsuki, The effects of elastic interaction energy on
  the $\gamma$? precipitate morphology of continuously cooled nickel-base
  alloys, Materials Science and Engineering 74~(2) (1985) 139--145.

\bibitem{fratzl1999modeling}
P.~Fratzl, O.~Penrose, J.~L. Lebowitz, Modeling of phase separation in alloys
  with coherent elastic misfit, Journal of Statistical Physics 95~(5-6) (1999)
  1429--1503.

\bibitem{karpov1998suppression}
S.~Y. Karpov, Suppression of phase separation in ingan due to elastic strain,
  Materials Research Society Internet Journal of Nitride Semiconductor Research
  3.

\bibitem{johnson2002mechanical}
B.~D. Johnson, B.~P. Boudreau, B.~S. Gardiner, R.~Maass, Mechanical response of
  sediments to bubble growth, Marine Geology 187~(3-4) (2002) 347--363.

\bibitem{algar2010stability}
C.~Algar, B.~Boudreau, Stability of bubbles in a linear elastic medium:
  Implications for bubble growth in marine sediments, Journal of Geophysical
  Research: Earth Surface 115~(F3).

\bibitem{liu2018methane}
L.~Liu, T.~De~Kock, J.~Wilkinson, V.~Cnudde, S.~Xiao, C.~Buchmann, D.~Uteau,
  S.~Peth, A.~Lorke, Methane bubble growth and migration in aquatic sediments
  observed by x-ray $\mu$ct, Environmental science \& technology 52~(4) (2018)
  2007--2015.

\bibitem{de1979effect}
P.-G. de~Gennes, Effect of cross-links on a mixture of polymers, Journal de
  Physique Lettres 40~(4) (1979) 69--72.

\bibitem{briber1988effect}
R.~M. Briber, B.~J. Bauer, Effect of crosslinks on the phase separation
  behavior of a miscible polymer blend, Macromolecules 21~(11) (1988)
  3296--3303.

\bibitem{style2018liquid}
R.~W. Style, T.~Sai, N.~Fanelli, M.~Ijavi, K.~Smith-Mannschott, Q.~Xu, L.~A.
  Wilen, E.~R. Dufresne, Liquid-liquid phase separation in an elastic network,
  Physical Review X 8~(1) (2018) 011028.

\bibitem{rosowski2020elastic}
K.~A. Rosowski, T.~Sai, E.~Vidal-Henriquez, D.~Zwicker, R.~W. Style, E.~R.
  Dufresne, Elastic ripening and inhibition of liquid--liquid phase separation,
  Nature Physics (2020) 1--4.

\bibitem{vidal2020theory}
E.~Vidal-Henriquez, D.~Zwicker, Theory of droplet ripening in stiffness
  gradients, arXiv preprint arXiv:2001.11752.

\bibitem{porter2009phase}
D.~A. Porter, K.~E. Easterling, M.~Sherif, Phase transformations in metals and
  alloys, (revised reprint), CRC press, 2009.

\bibitem{mooney1940theory}
M.~Mooney, A theory of large elastic deformation, Journal of applied physics
  11~(9) (1940) 582--592.

\bibitem{rivlin1948large}
R.~Rivlin, Large elastic deformations of isotropic materials. i. fundamental
  concepts, Philosophical Transactions of the Royal Society of London. Series
  A, Mathematical and Physical Sciences 240~(822) (1948) 459--490.

\bibitem{raayai2019volume}
S.~Raayai-Ardakani, Z.~Chen, D.~R. Earl, T.~Cohen, Volume-controlled cavity
  expansion for probing of local elastic properties in soft materials, Soft
  matter 15~(3) (2019) 381--392.

\bibitem{alan2001engineering}
G.~Alan, Engineering with rubber: how to design rubber components (2001).

\bibitem{gent1969surface}
A.~Gent, D.~Tompkins, Surface energy effects for small holes or particles in
  elastomers, Journal of Polymer Science Part A-2: Polymer Physics 7~(9) (1969)
  1483--1487.

\bibitem{rosowski2020elastic2}
K.~A. Rosowski, E.~Vidal-Henriquez, D.~Zwicker, R.~W. Style, E.~R. Dufresne,
  Elastic stresses reverse ostwald ripening, arXiv preprint arXiv:2004.05070.

\end{thebibliography}
\end{document}